\documentclass{elsart}
\journal{New Astronomy}
\usepackage{natbib}
\usepackage{graphicx,epsfig}
\usepackage{amssymb}

\begin{document}

\begin{frontmatter}

\title{Detecting the Transition \\ From Pop III to Pop II Stars}

\author{Aparna Venkatesan}
\address{NSF Astronomy and Astrophysics Postdoctoral Fellow, \\ Center for
  Astrophysics and Space Astronomy, UCB 389, \\ University of Colorado, Boulder, CO 80309-0389  \\
E-mail: aparna.venkatesan@colorado.edu}

\begin{abstract}

We discuss the cosmological significance of the transition from the Pop III
to Pop II mode of star formation in the early universe, and when and how it
may occur in primordial galaxies. Observations that could detect this
transition include those of element abundances in metal-poor Galactic halo
stars, and of the helium reionization and associated heating of the
intergalactic medium. We suggest that gamma-ray bursts may be a better
probe of the end of the first-stars epoch than of Pop III stars.

\end{abstract}

\begin{keyword}
cosmology \sep theory \sep Population III stars \sep cosmic microwave
background \sep reionization \sep intergalactic medium \sep gamma ray burst
\end{keyword}

\end{frontmatter}

\section{The End of the First Stars Epoch: A Cosmic Milestone}

The nature of the primordial stellar initial mass function (IMF) is
currently a problem of great interest.  Some recent theoretical work
indicates that this IMF may have been top-heavy \citep{abeletal00,
bromm02}, leading predominantly to stellar masses $\geq$ 100 $M_\odot$ up
to a critical gas metallicity, $Z_{\rm cr} \approx 10^{-4 \pm 1} Z_\odot$
\citep{brommetal01, schneider02}, above which a present-day IMF
occurs. However, other detailed studies of stellar feedback on accreting
matter \citep{tanmckee} and of the current data on reionization and the metal
abundance ratios in extremely metal-poor (EMP) Galactic halo stars
\citep{tvs04} suggest otherwise: that the primordial IMF, rather
than being biased towards high masses, may merely lack low-mass stars.  The
transition metallicity $Z_{\rm cr}$ may be significantly higher at low
densities and could vary with metal species such as C, O and Fe which are
important coolants in metal-free gas \citep{brommloeb03, santoroshull}.

Stars of masses $\sim$ 10--100 $M_\odot$ end their lives as the more
familiar Type II supernovae (SNe), leaving behind neutron stars and black
holes, whereas metal-free stars in the mass range $\sim$ 140--260 $M_\odot$
are thought to disrupt themselves entirely as pair-instability SNe (PISNe).
A first-stars metal-free IMF spanning stellar masses 10--140 $M_\odot$ is
fully consistent with the current data on reionization -- including the
enhanced electron-scattering optical depth, $\tau_e \sim$ 10--15\% in the
cosmic microwave background (CMB) -- and on nucleosynthetic abundances in
EMP stars \citep{tvs04}, the intergalactic medium (IGM;
\citealt{venktruran}), and in QSO broad-emission-line regions at high
redshifts, $z$ \citep{vsf04}. This requires metal-free star formation (SF)
to last for $\sim 10^7$--$10^8$ yr in early galactic halos, which is
entirely consistent with calculations on the duration of the Pop III epoch
from semi-analytic \citep{tvs04} and numerical methods \citep{wadavenk,
byh03}. Since the transport of metals and radiation in primordial galaxies
and the IGM is almost certain to be highly inhomogeneous, the transition
from a Pop III (metal-free first stars) to Pop II (second-generation
metal-poor) IMF, illustrated in Figure 1, may vary significantly in space
and time.  The onset of Pop II SF may be non-unique cosmologically, rather
than an abrupt global transition.

\begin{figure}[!t]
\centerline{\psfig{file=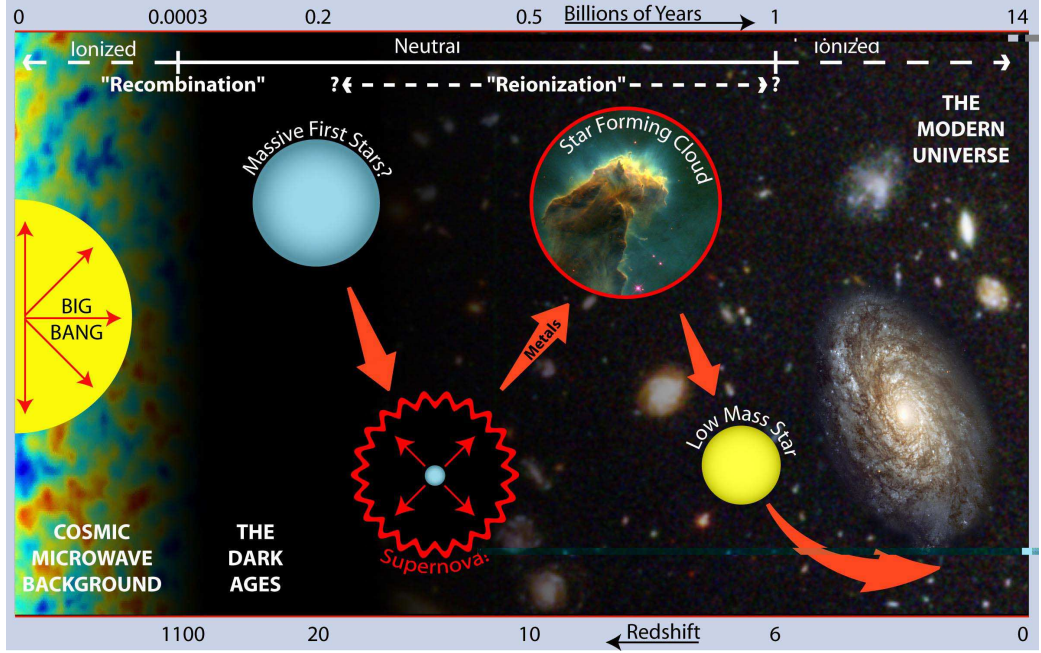,width=5.4in}}
\caption{A schematic representation of the transition at early cosmological
  epochs from the first metal-free stars to second-generation star
  formation, following the generation of metals from the first SNe. The H
  and He reionization of the IGM may be completed by early stellar
  generations prior to the epochs of massive QSOs at $z < 6$, with partial
  or complete recombination during the metallicity transition
  phase. In this case, a second reionization of H at $z \sim 6$ and
  of He at $z \sim 3$ subsequently occur.}
\end{figure}

Regardless of the uncertainties related to the primordial stellar IMF and
the transition epochs to Pop II SF, we expect the first generations of
stars to form from metal-free gas. Their composition heavily influences
their structure and properties, as they rely predominantly on the p--p
chain initially than on the more efficient CNO cycle for their
thermonuclear fuel source \citep{tsv}. Consequently, metal-free stars are
hotter and emit significantly harder ionizing radiation relative to their
finite-$Z$ counterparts \citep{bkl, tsv, sch02}.  Thus, the
traditional distinction between the hard ionizing spectra of QSOs and the
relatively soft spectra of stars fails in the case of Pop III SF. This
has critical consequences for several problems in cosmology, as we detail
below.

Last, we note that gas transition metallicities of $Z_{\rm cr} \sim
10^{-3.5} Z_\odot$ mark a genuine milestone in cosmic history. Using
current stellar evolution models, we can derive that the minimum number of
ionizing photons per baryon that must have been generated in association
with this value of $Z_{\rm cr}$ (whether in total $Z$, or individual
metals) is at least 1--10 \citep{venktruran}. This is because metal free
stars in a present day IMF are 10--20 times more efficient at generating
ionizing radiation per metal yield than are solar-$Z$ stars \citep{sch02,
venktruran}. Therefore, the universe may well have been reionized by the
time $Z_{\rm cr}$ is achieved in individual halos or the IGM, an important
potential feature of the cosmic backdrop in which Pop II SF commences.

\section{Constraints from the CMB}

The 1-year temperature-polarization correlation data of the CMB from $WMAP$
implies a large Thomson optical depth of $\tau_e \sim 0.17 \pm 0.08$
\citep{spergel03}. In a single-step reionization scenario, where the IGM
abruptly transitions from being neutral to completely ionized, this
translates to a reionization epoch, $z_{\rm reion} \sim 17 \pm 4$. At first
glance, this is not compatible with Gunn-Peterson (GP) studies of the
high-$z$ IGM, which indicate $z_{\rm reion} \sim 6$ (\citealt{vts} and
references therein). This raises the possibility of multiple reionizations
or extended periods of partial ionization of H and/or He over $z \sim$
6--20 \citep{vts, cen03}. As we emphasized earlier, a non-exotic stellar
IMF consisting of metal-free stars in the mass range 10--140 $M_\odot$ can
lead to values of $\tau_e \sim$ 0.1--0.15 \citep{tvs04}.  We must also
consider the additional contributions from X-rays from the first stars and
QSOs \citep{vgs, ricottiost04} and from He reionization, each of which may
contribute a few percent. One possible method to disentangle these two
differing contributions to $\tau_e$ in the CMB is to utilize future data at
radio wavelengths in combination with the CMB to constrain the topology of
reionization (see, e.g., S. Furlanetto, this proceedings). The far-UV
He-ionizing photons would be stopped close to the halos in sharp I-fronts
which slowly advance into the IGM, whereas X-ray ionization would resemble
fuzzy halos around the host galaxy, owing to the high penetrating power of
X-rays.

We note that GP and CMB constraints are not necessarily incompatible when
we consider that these techniques are sensitive to complementary stages of
reionization. The first best probes trace amounts of neutral matter in the
IGM whereas CMB photons experience the most scattering with IGM electrons
during a sharp rise of IGM ionization. Furthermore, $\tau_e$ is only a
cumulative measure whose values can be reproduced by a number of
reionization histories. We suggest therefore that future measurements of
CMB polarization, from $WMAP$ or ground-based experiments, in combination
with direct imaging of primordial stellar clusters may provide the best use
of CMB data to constrain the transition from Pop III to Pop II SF
at high redshifts.

\section{The Role of Helium}

\begin{figure}[!t]
\centerline{\psfig{file=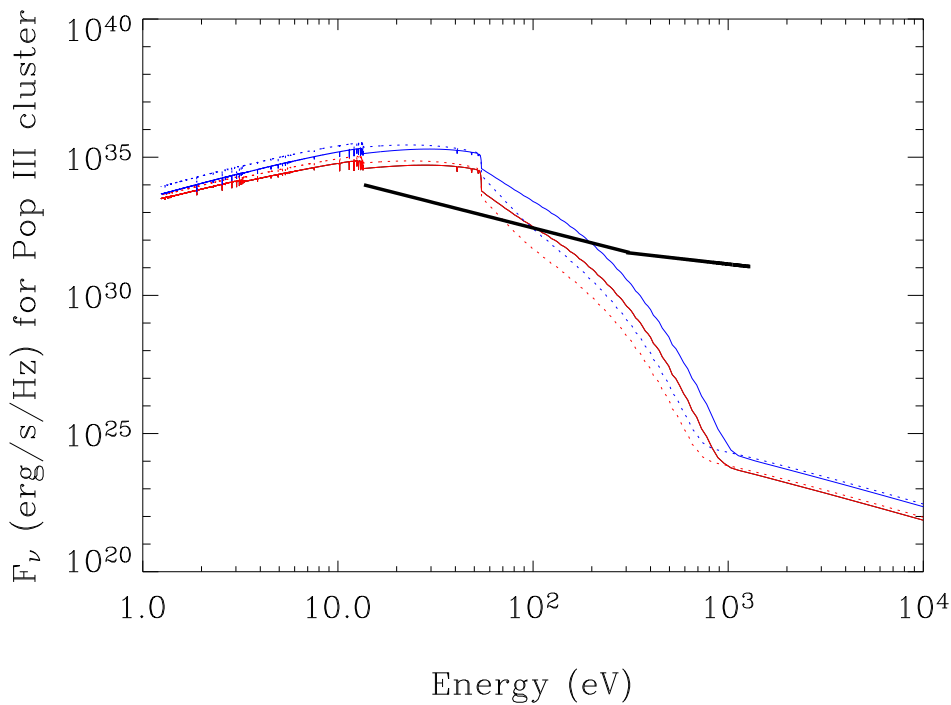,width=5.4in}}
\caption{The specific intensity as a function of energy of a fiducial
  $10^6$ $M_\odot$ metal-free stellar cluster, based on Tumlinson et al. (2004). Solid and dashed lines denote the values on the ZAMS and at times of 2
  Myr; within each of these cases, upper and lower lines represent
  1--100 $M_\odot$ and 10--140 $M_\odot$ IMFs.  Note the
  relative flatness (hardness) of the cluster's radiation between the H and He
  ionizing thresholds at 13.6 eV and 54.4 eV respectively, and the rapid
  decline of the He ionizing flux with time. For comparison, a typical QSO
  spectrum (arbitrarily normalized) with slopes 1.8 and 0.8 in the energy
  ranges 13.6--300 eV and $\geq$ 300 eV is also shown.}
\end{figure}

As emphasized earlier, the first metal-free stars are significantly hotter
and emit harder ionizing radiation than do low-$Z$ stars of the same
mass. Specifically, a $Z=0$ stellar cluster creates 60\% more H-ionizing
photons and $10^5$ more He-ionizing photons than a Pop II $Z=0.001$ cluster
in a Salpeter IMF over 1--100 $M_\odot$ \citep{vts}. This could lead to one
or more reionization epochs for both H and He, boosting the IGM electron
fraction and the detected $\tau_e$ in the CMB. The Thomson optical depth
depends on the line-of-sight integral of the number density of electrons,
$n_e$, which for a fully reionized IGM becomes, $n_e$ = $n_{\rm H}$ +
2$n_{\rm He}$. For $\frac{n_{\rm He}}{n_{\rm H}} = 0.08$, we see that
complete He reionization can increase $\tau_e$ in the CMB up to an extra
16\% of its value from H reionization alone, a small but significant
contribution. The end of Pop III SF will lead to a loss of He-ionizing
radiation and likely cause recombination of He~III in the IGM. The
accompanying emission of 4 Ryd photons and the consequent heating of the
IGM may be detected through the thermal evolution of the IGM down to $z
\sim 3$ \citep{huihaiman}. Thus, the onset of Pop II star formation could
be tracked through the He reionization and thermal history of the IGM.

An important factor in determining the evolution of H and He I-fronts and
the geometry of ionized bubbles in primordial galaxies is the frequency
dependence of the source spectrum.  For a source whose specific intensity
(erg s$^{-1}$ Hz$^{-1}$) goes as $F_\nu \propto \nu^{-\alpha}$, we find
that the ratio of photoionization rates per atom for He and H, $R$, is : \\
\begin{equation}
R \equiv \frac{Q({\rm HeII})/n_{\rm He}}{Q({\rm HI})/n_{\rm H}} =
\left[ \frac{\int_{\rm 4 Ryd}^\infty d\nu F_\nu/(h \nu)}{\int_{\rm 1
      Ryd}^\infty d\nu
  F_\nu/(h \nu)} \right] \frac{n_{\rm H}}{n_{\rm He}}
\end{equation}

For $\Omega_b h^2$ = 0.0224 and $y = n_{\rm He}/n_{\rm H}$ = 0.08, the
above ratio is unity when $\alpha_{\rm crit} \sim 1.82$.  This is the
critical spectral index at which the rate of H and He photoionizations
become equal; for a sufficiently hard source, the He and H I-fronts may
begin to coincide. Current calculations \citep{venkshull} indicate that Pop
III stars in a Salpeter slope 1--100 $M_\odot$ or 10--140 $M_\odot$ IMF may
be unable to achieve this criterion ($R \geq 1$), especially when
evolutionary effects are taken into account. This is shown in Figure 2,
where we display the specific intensity of a fiducial $10^6 M_\odot$ Pop
III cluster as a function of energy for these two IMFs.  A numerical
evaluation of $R$ for these cases reveals that it never exceeds unity,
mostly due to the sharp falloff of the intensity beyond the He~II
ionization threshold. However, a halo that hosts a QSO (also shown in
Figure 2) in addition to Pop III stars could well result in He and H
I-fronts that advance together spatially.

As noted in \cite{vts}, future data from, e.g., the {\it Cosmic Origins
Spectrograph} on the {\it Hubble Space Telescope}, on the He~II GP effect
at $z >3$ can place limits on relic ionization from Pop III stars,
particularly in underdense regions of the IGM that may not have recombined
after the Pop III era.  Such data can also constrain the topology of
escaping He-ionizing radiation and the relative roles of ionization from
hard (Pop III stars and QSOs) and soft (Pop II stars) sources.

\section{Dust Transport and Pop II Metallicities}

Studies of EMP stars in the Galactic halo provide an excellent avenue
complementary to those detailed above for probing the transition from Pop
III to Pop II SF.  Surviving members of low-mass second-generation stellar
populations that form in the cooling shells of Pop III SN remnants (SNRs)
may be detected as EMP stars at present.  A large number of elements have
been detected in many EMP stars, and their relative abundance ratios
provide strong constraints on the IMF, formation conditions and chemical
environment of early stellar generations. We present here some recent
results on the role of dust transport in the segregated metal enhancement of
Pop II starforming sites, and refer the interested reader to \cite{tvs04}
for a detailed analysis of the current nucleosynthetic data on EMP stars
and related limits on the first-stars IMF.

\begin{figure}[!t]
\centerline{\psfig{file=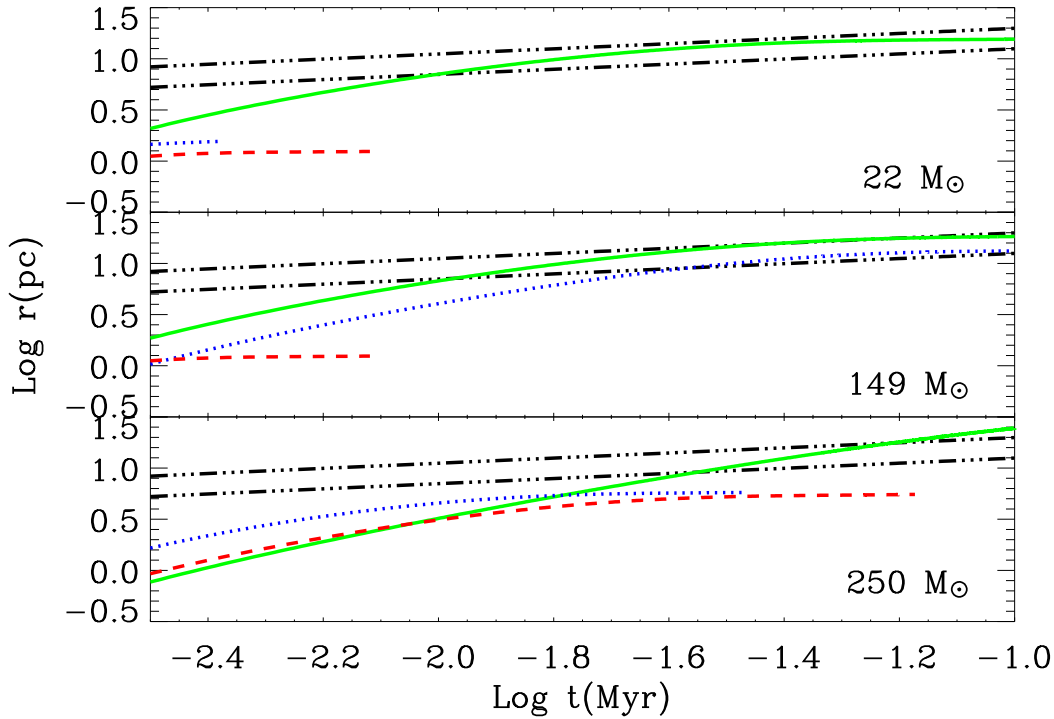,width=5.4in}}
\caption{The positions of dust grains within SNRs in primordial galaxies
are shown as a function of time as they are radiatively transported outward
from the center of the SNR. Top, middle and bottom panels represent stellar
masses of 22 $M_\odot$, 149 $M_\odot$, and 250 $M_\odot$. Solid, dotted and
dashed curves refer to graphites, silicates and magnetite grains.  The two
long dashed lines in each panel describe the adiabatic evolution of the SN
shell radius, for $E_{\rm SN} = 10^{52}$ erg and and ambient gas number
densities of 1 cm$^{-3}$ (upper line) and 10 cm$^{-3}$ (lower line).}
\end{figure}

In \citet{vns05}, we investigated the radiative transport of dust within
SNRs in primordial galaxies, motivated by the strong overabundances of the
elements Mg, Si, O and C at low iron abundances in EMP stars. These trends
are not easily explained by energetic Type II SN models alone. We consider
a dust transport scenario involving graphite, silicate and iron-bearing
magnetite grains, the dominant products of PISNe and metal-free Type II
SNe.  We include the effects of radiation pressure from Pop III stars,
gravity, gas drag, grain charging, and sputtering on grain dynamics, and
require only Pop III cluster luminosities of $\sim 10^6 L_\odot$ over
timescales of $10^5$ yr, and SN kinetic energies of at least $10^{51}$ erg.
We find, for reasonable parameter assumptions, that the transport of the
primary dust compounds within early SNRs can be segregated, leading to a
differential accumulation of metals that seed Pop II SF in SN shells.  The
results from \citet{vns05} can at least partly account for the element trends
observed in C- and silicate-rich Fe-poor EMP stars, whose abundances are
best explained in our models by a Pop III IMF over masses $\sim$ 10--150
$M_\odot$, as displayed in Figure 3.

\section{Gamma-Ray Bursts From Pop III Stars?}

In concluding our discussion of detecting Pop III and Pop II stars at high
redshifts, we briefly examine the viability of constraining the first-stars
epoch through gamma-ray bursts (GRBs). Many authors have proposed that GRBs
associated with Pop III stars could be the most effective way to probe
high-$z$ SF, IGM ionization and metallicity at $z \geq 6$ (see, e.g.,
D. Lamb, this proceedings). At these redshifts, emission lines from QSOs
and galaxies will become increasingly hard to detect owing to natural
dimming with distance and absorption by the neutral IGM, whereas GRB
afterglows experience a fortuitous near-constancy of spectral flux in an
observed frequency range with increasing redshift. The association of GRBs
with metal-free stars is largely theoretical at present; the observational
evidence rests primarily on the GRB-SN association at low redshifts. The
leading models for the long- and short-duration GRBs are respectively
\citep{hurley} the single progenitor model, involving the core collapse of
a massive star through a SN event, and binary models involving neutron
stars and black holes. Let us examine these in turn.

First, the massive star model has been tested through the low-$z$
detections of GRBs in association with metal-poor SNe of energies $\sim
10^{51}$--$10^{53}$ erg. However, these SNe are also observed to be Type
Ib/c in nature, i.e., lacking in hydrogen and/or helium envelopes. The Pop
II progenitors of these objects have likely experienced significant mass
loss, a critical factor for the success of the GRB mechanism. We may then
question whether extending such low-$z$ analogues to Pop III stars is
appropriate or realistic, and whether the GRB engine can operate in the
core of a compact Pop III star. After all, metal-free stars are essentially
composed of a H/He envelope in their entirety!  Second, the shortest
timescales on which a short-duration GRB may form is of order 10 Myr from a
high-mass binary system. We recall that on this timescale, an individual
halo may well have experienced metal self-pollution that is sufficient to
halt further Pop III SF. A GRB created in a binary scenario may then
provide excellent constraints on the transition metallicity at which
second-generation SF occurs in high-$z$ halos. This problem clearly
deserves more detailed investigation as well as more data on the GRB-SN
connection. For now, we tentatively suggest that GRBs may in reality be a
better probe of the {\it end} of the first-stars phase and the onset of Pop
II SF.

\section{Discussion}

We have presented recent work that places limits on the IMF, epochs and
formation conditions of Pop III and Pop II SF from a variety of data.  The
ideal method to test the predictions of these calculations would clearly be
through direct detections of Pop III and/or Pop II stellar clusters at $z
\geq 6$, through emission-line (He~II, Ly$\alpha$) and color signatures
(\citealt{tsv}, Rhoads \& Malhotra, this proceedings). The viability of
such detections depends strongly on the converse problem to the impact of
Pop III stars on reionization, which is the effective trapping of radiation
within primordial galaxies. A related theoretical goal is to better
understand the duration of Pop III SF and the chemical feedback from the
first SNe in driving the transition to Pop II SF: how quickly is the
environment to form Pop III stars lost and is this a highly nonuniform
process spatially and temporally?  What are the timescales for halo
self-enrichment versus pollution of neighboring galaxies? What are the
metallicities of second-generation stars forming in the wake of the first
SNe?  These questions have a direct impact on the cosmological relevance of
metal-free stars and on the planning of future missions that are explicitly
designed to detect ``First Light'' such as the {\it James Webb Space
Telescope}. We have attempted here to address some of these issues. Further
theoretical progress in this field may depend on numerical simulations
which are best suited to study problems such as these, whose nature
inherently involves inhomogeneity and complex local feedback processes.

Other techniques to constrain the Pop III epochs and the transition to Pop
II SF include the detection of GRBs, and of the near-IR and radio signals
from early stellar sources as discussed by other authors in this
volume. The discovery of a metal-free star in the local universe would also
be an important stride towards understanding Pop III SF. This is however an
exceedingly difficult observation in practice -- as we mentioned in our
talk, detecting the most iron-poor star currently known, a 1 part in $10^9$
detection, is already equivalent to finding the author of this contribution
in all of India!  Until direct detections of a truly metal-free star become
possible, indirect inferences must be made from the IGM reionization of H
{\it and} He, the metal content of high-$z$ systems, and stellar fossils in
our Galactic backyard. Of these, EMP stars which could be survivors of the
transition epochs from Pop III to II SF are likely the most fruitful avenue
to constrain this problem in the near future.

I thank my collaborators Jason Tumlinson, Mike Shull, Jim Truran, Biman
Nath, Andrea Ferrara, Raffaella Schneider and Keiichi Wada for the projects
and conversations that led to the results presented here. I also thank
Jason Tumlinson for helpful correspondence and for sharing Figure 1 which
he created for press releases on our joint work in 2004. I gratefully
acknowledge the support of NSF grant AST-0201670 through the NSF Astronomy
and Astrophysics Postdoctoral Fellowship program.

\end{document}